\documentclass{article}

\usepackage{microtype}
\usepackage{graphicx}
\usepackage{subfigure}
\usepackage{booktabs} %

\usepackage{xspace}
\usepackage{xcolor}
\usepackage{times}
\usepackage{latexsym}
\usepackage{graphicx}
\usepackage{url}
\usepackage{lipsum}
\usepackage{amsmath}
\usepackage{amssymb}
\usepackage{multirow}
\usepackage{arydshln}
\newcommand{\ie}{i.e.,\xspace}

\newcommand{\eat}[1]{}

\usepackage{hyperref}

\usepackage[accepted]{icml2023}

\icmltitlerunning{LongCoder: A Long-Range Pre-trained Language Model for Code Completion}

\begin{document}

\twocolumn[
\icmltitle{LongCoder: A Long-Range Pre-trained Language Model for Code Completion}

\icmlsetsymbol{equal}{*}

\begin{icmlauthorlist}
\icmlauthor{Daya Guo}{equal,to}
\icmlauthor{Canwen Xu}{equal,goo}
\icmlauthor{Nan Duan}{ed}
\icmlauthor{Jian Yin}{to}
\icmlauthor{Julian McAuley}{goo}
\end{icmlauthorlist}

\icmlaffiliation{to}{Sun Yat-sen University}
\icmlaffiliation{goo}{University of California, San Diego}
\icmlaffiliation{ed}{Microsoft Research Asia}

\icmlcorrespondingauthor{Daya Guo}{guody5@mail2.sysu.edu.cn}

\icmlkeywords{Language Model, Code Completion, Code Generation, Sparse Transformer, Long Transformer}

\vskip 0.3in
]

\printAffiliationsAndNotice{\icmlEqualContribution} %

\begin{abstract}

In this paper, we introduce a new task for code completion that focuses on handling long code input and propose a sparse Transformer model, called LongCoder, to address this task. LongCoder employs a sliding window mechanism for self-attention and introduces two types of globally accessible tokens --- \textit{bridge tokens} and \textit{memory tokens} --- to improve performance and efficiency. \textit{Bridge tokens} are inserted throughout the input sequence to aggregate local information and facilitate global interaction, while \textit{memory tokens} are included to highlight important statements that may be invoked later and need to be memorized, such as package imports and definitions of classes, functions, or structures. We conduct experiments on a newly constructed dataset that contains longer code context and the publicly available CodeXGLUE benchmark. Experimental results demonstrate that LongCoder achieves superior performance on  code completion tasks compared to previous models while maintaining comparable efficiency in terms of computational resources during inference.
\end{abstract}

\section{Introduction}

\begin{figure*}[t!]
    \centering
    \includegraphics[width=0.95\linewidth]{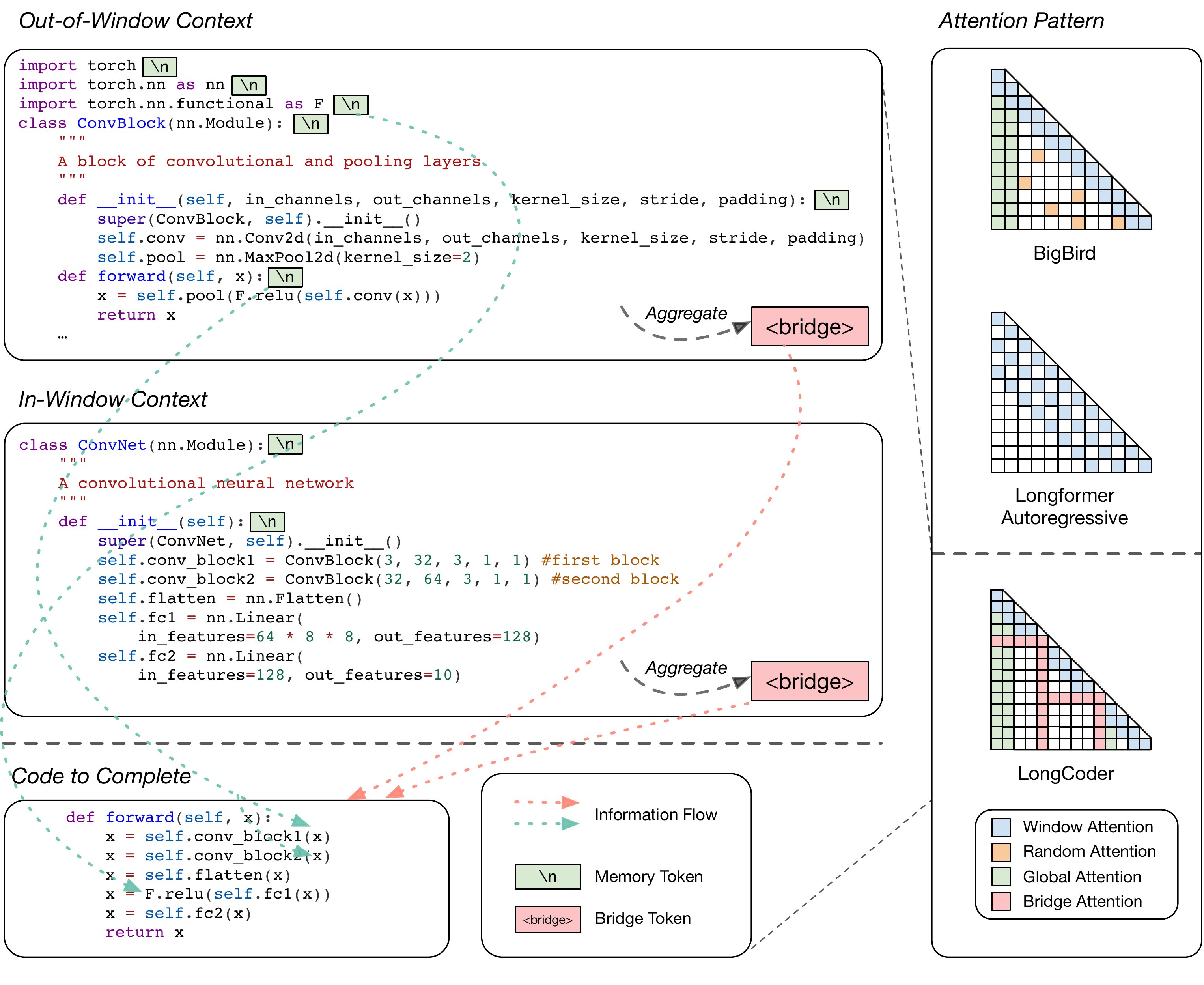}
    \caption{\textit{\textbf{(Left)}} An example of how LongCoder facilitates completion with longer context. The memory tokens save potentially useful information (including package imports, class and function definitions) for global access despite whether they are within the sliding window. The bridge tokens aggregate local information by attending to a fixed length of tokens. The information flow within the window is omitted for clarity.
    \textit{\textbf{(Right)}} Attention patterns used in BigBird~\cite{bigbird}, Longformer~\citep{beltagy2020longformer} and LongCoder. Best viewed in color.}
    \label{fig:code}
\end{figure*}

Code completion is a crucial task in software development that helps developers save time and effort by suggesting and auto-completing code based on context. With the advancement of large language models, Transformer-based models~\citep{transformer} have demonstrated impressive results in code completion~\citep{chen2021evaluating}. However, the computational cost of these models grows quadratically with the length of input, making them less suitable for modeling long code context. On the other hand, modeling long code can potentially improve the accuracy of code completion and enable applications on a file and even project level. An efficient model that can scale to such long input can be suitable for code completion that contains long context. 

In this paper, we propose a new pre-trained language model, named LongCoder, for long code modeling. As shown in Figure~\ref{fig:code}, LongCoder features a sparse attention mechanism that reduces the computational complexity (to linear). LongCoder exploits a sliding window mechanism for self-attention that attends only to local context. To allow LongCoder to maintain an understanding of the entire code file, we introduce \textit{bridge attention} and \textit{global attention}, with the corresponding two types of globally accessible tokens, \textit{bridge tokens} and \textit{memory tokens}. Bridge attention aggregates the information of a code snippet and allows it to be accessed from a long distance. Bridge tokens are inserted throughout the input sequence and can attend to a fixed length of context. Memory tokens provide global attention to statements that include a package import, definitions of classes, functions, or structures. %
The scope of these statements is often global and invoked later, which means they have a longer impact than other statements, making them worth memorizing. By referring to these statements, the model can exploit long context while maintaining %
linear complexity.

To evaluate the effectiveness of LongCoder and encourage future research on \textbf{L}ong \textbf{C}ode \textbf{C}ompletion, we construct a new dataset called LCC by filtering code from GitHub based on length, with the goal of focusing on longer code examples. On average, the examples in LCC are 5$\times$ longer than those in existing datasets~\citep{lu2021codexglue}. We benchmark several baselines, LongCoder and OpenAI Codex~\citep{chen2021evaluating} on LCC. Our experimental results demonstrate that code completion can benefit from taking longer context into consideration, and our LongCoder achieves superior performance compared to existing models with comparable %
computational costs.

Overall, our contributions are as follows:
\begin{itemize}
    \item We construct a new dataset (LCC) for code completion tasks that requires long code modeling to encourage more research in such scenarios.
    \item We propose two types of sparse attention, motivated by observations on attention patterns of existing models and how human programmers write code.
    \item We train and release LongCoder, a sparse and efficient pre-trained Transformer model for long code modeling, which achieves superior performance on both long and regular code completion with comparable computational resources.\footnote{All the codes and data are available at \url{https://github.com/microsoft/CodeBERT}.}  
\end{itemize}

\section{Related Work}
\paragraph{Code Completion}
Code completion is an essential task that helps programmers improve their efficiency by suggesting and automatically completing code based on context and previous inputs. Prior works have explored the use of statistical learning for the code completion task, such as the use of n-gram techniques \cite{tu2014localness,hindle2016naturalness} and probabilistic grammar-based methods \cite{allamanis2014mining,bielik2016phog,
raychev2016probabilistic,hellendoorn2017deep}. With the  success of pre-training in natural language processing \cite{devlin2018bert,radford2019language,brown2020language},  decoder-only pre-trained models based on Transformer have been proposed to promote the development of code completion.  \citet{svyatkovskiy2020intellicode} and \citet{lu2021codexglue} respectively propose GPT-C and CodeGPT, which are pre-trained by generating code from left to right in an auto-regressive manner on large amounts of code. \citet{liu2020multi} and \citet{guo2022unixcoder} pre-train similar models CugLM and UniXcoder with multi-task learning by leveraging code structure for code completion. Codex \cite{chen2021evaluating}, PolyCoder \cite{xu2022systematic}, CodeGen \cite{nijkamp2022conversational}, InCoder \cite{fried2022incoder}, and AlphaCode \cite{li2022competition} build large language models with 
billions of
parameters and achieve impressive performance on code generation by training on a large-scale and high-quality code corpus. For these pre-trained models, it is impractical to simply expand the context window to model long-range sequences, due to computational complexity of the attention mechanism increasing quadratically with the input length. Therefore, \citet{clement2021long} propose to extract the most important code fragments and integrate them into a fixed-length context window. However, due to the constraint of fixed window length, some high-priority code, such as class and function definitions, may be omitted. Additionally, increasing the window length would also introduce additional computational overhead. Different from these works,  LongCoder is a sparse Transformer that can take advantage of the entire file-level code context while maintaining comparable efficiency in terms of computational resources during inference.

\paragraph{Long-Range Transformer Models} The original Transformer~\citep{transformer} is inefficient for modeling long sequences since its time and space complexity is $O(n^2)$, where $n$ is the length of the sequence. Prior studies focus on optimizing the complexity to enable processing of longer sequences. To name a few, Sparse Transformer~\citep{child2019generating} reduces the quadratic complexity of standard self-attention  by computing attention on sparse query-key pairs. Sparse Transformer uses a dilated sliding window to capture  local context. Reformer~\citep{reformer} proposes locality sensitive hashing (LSH) attention to reduce the complexity and memory footprint. Longformer~\cite{beltagy2020longformer} uses dilated sliding windows to model longer sequences and adds global memory tokens to allow interaction with all tokens. Performer~\citep{performer} generalizes attention calculation by introducing kernel functions. They then propose a random kernel function, namely orthogonal random features (ORF) to approximate the standard self-attention. Linformer~\cite{wang2020linformer} applies low-rank projection to the length dimension to reduce the complexity of self-attention. Linear Transformers~\cite{lineartransformer} uses a kernel function that exploits the associativity property of matrix products to reduce  complexity. 
BigBird~\citep{bigbird} has an attention pattern comprised of random attention, window attention and global attention. CosFormer~\citep{cosformer} proposes a linear operator and a cosine-based distance re-weighting mechanism as the substitute for softmax attention.
We recommend \citet{tay2022efficient} as a more comprehensive survey on long-range efficient Transformer models. Different from these works, our LongCoder introduces code heuristics into the dynamic construction of global attention to imitate how human programmers code.

\section{Long Code Completion}
\label{sec:lcc}

Code completion is a fundamental and important task for code models, %
which can help programmers improve their efficiency while coding. Previous public benchmarks primarily focused on completion with short code context. For instance, CodeXGLUE \cite{lu2021codexglue} offers two code completion datasets from \textit{PY150} \cite{raychev2016probabilistic} in Python and \textit{Github Java Corpus} \citet{allamanis2013mining} in Java, and also builds two test datasets to evaluate next-line prediction. The average length of the code context in the two test datasets is 478 tokens and 365 tokens, respectively. However, according to our statistics, the average length of a Python source file on GitHub is 1,305 tokens. After tokenization, %
the average length becomes 2,090 tokens while 41\%/24\% of the files have a length longer than 1,024/2,048 tokens, which highlights the need for models that can handle longer code sequences in order to be more practical and useful in the real-world.
Meanwhile, longer code sequences contain more complex structures and require models to consider more context and dependencies. This can be challenging for previously proposed code completion models that focus on short code and do not take into account the long context of the code. By evaluating models on longer code sequences, we can better understand their ability to handle more complex and realistic scenarios. Meanwhile, long code completion poses new challenges for efficiency of code models, as in vanilla Transformers~\citep{transformer}, the computational resources grow quadratically with the input length.

In this paper, we introduce the \textbf{L}ong \textbf{C}ode \textbf{C}ompletion Benchmark (\textbf{LCC}), a new benchmark that focuses on code completion with long code context for three programming languages: Python, Java, and C\#. Specifically, we construct our datasets from the \textit{github-code}\footnote{\url{https://huggingface.co/datasets/codeparrot/github-code}} dataset, which contains a vast number of code files sourced from GitHub with an open-source license that permits research use. The steps to construct the datasets are as follows:
\begin{itemize}
\item We first follow \citet{allamanis2019adverse} to deduplicate examples with high similarity (Jacobi similarity $\ge$ 0.9) in order to eliminate forked files, and then remove code files that can't be parsed into an abstract syntax tree using a standard compiler tool called \textit{tree-sitter}.\footnote{\url{https://github.com/tree-sitter/tree-sitter}}

\item Since the benchmark primarily focuses on the code completion task with long code context, we remove code files whose length of code tokens after tokenization is shorter than 512. Additionally, we also eliminate excessively long code files with a length greater than 10,000 tokens.

\item For each programming language, we sample 100k examples for training, and 10k examples for development and 10k for testing.
For each sample on development and test sets, we randomly sample an uncommented line of code not shorted than 3 tokens and ensure that there is sufficient context, i.e., a context larger than 512 code tokens. The data statistics of the context length in the LCC test sets are listed in Table~\ref{LCC-sta}. 
\end{itemize}
 \begin{table}[t]
\caption{Data statistics of the code context length in LCC test set. 25\%/50\%/75\% refer to the first/second/third quartile.}
\label{LCC-sta}
\begin{center}
\begin{tabular}{lcccc}
    \toprule
    Language&Average &25\% &50\% &75\% \\
    \midrule
    Python &1993.3& 1056& 1438& 2211\\
    Java  &1841.4& 1058& 1307& 2003\\
    C\#  &1970.5& 1023& 1396& 2143\\
    \bottomrule
\end{tabular}
\end{center}
\end{table}

We follow \citet{lu2021codexglue} to evaluate the performance of the models in terms of Exact Match (EM) and Edit Similarity (Edit Sim) on a per-line basis~\citep{svyatkovskiy2020intellicode}.

\section{LongCoder}
LongCoder is an attempt to tackle the efficiency problem of modeling longer code. It applies sparse attention to reduce quadratic time and space complexity of self-attention to linear. There are three types of attention in LongCoder --- window attention, bridge attention, and global attention. Each type is motivated by observations on previous models and focuses on one important aspect in modeling long code. The three types of attention are illustrated in Figure~\ref{fig:code} and we will describe them individually.

\subsection{Window Attention}
\begin{figure}[t!]
    \centering
    \includegraphics[width=0.95\columnwidth]{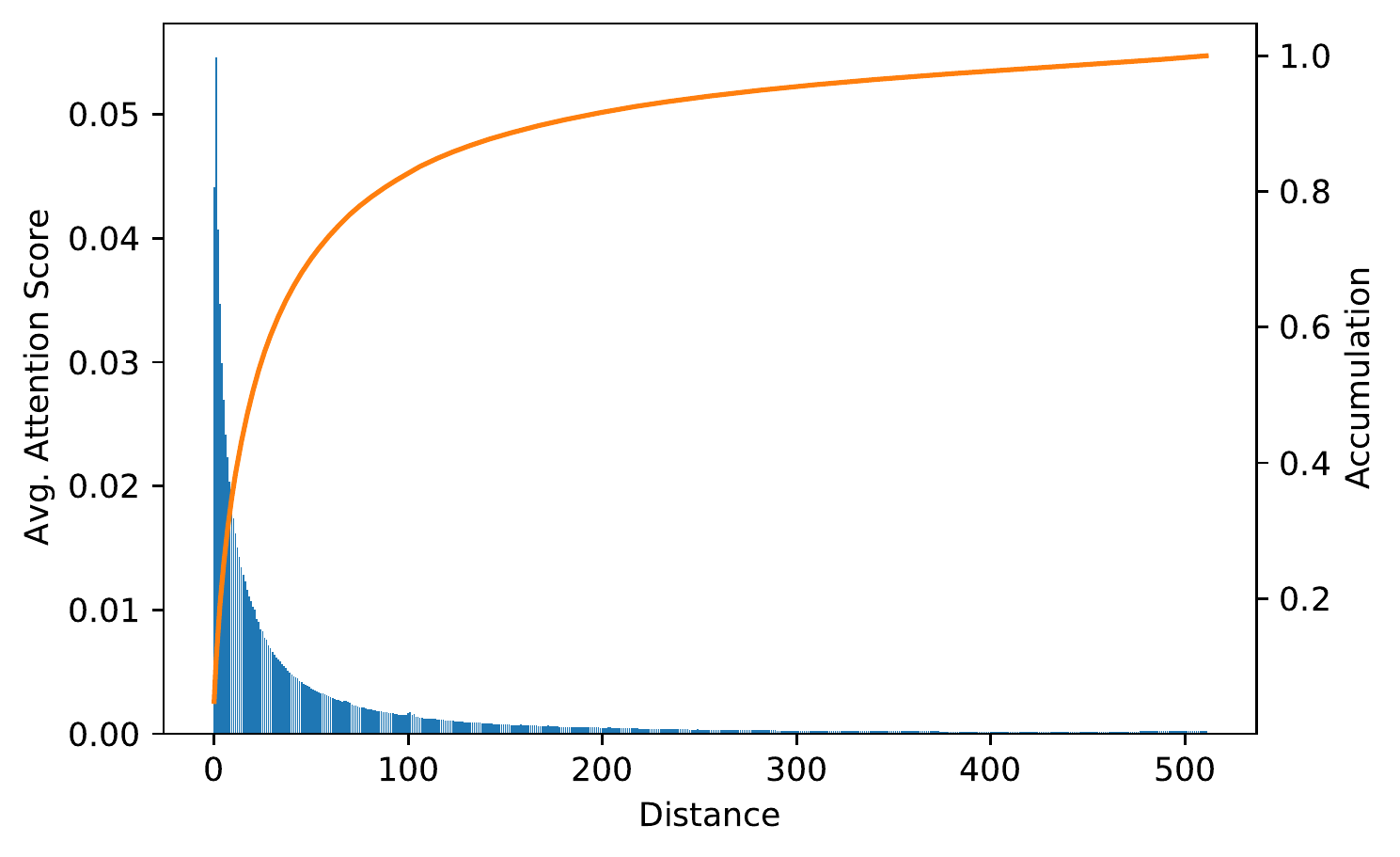}
    \caption{Distribution of average attention scores between two tokens within different distances in CodeGPT~\citep{lu2021codexglue}. The attention score is an average of 100 Python code examples across all Transformer layers.}
    \label{fig:window_attn}
\end{figure}

Code completion largely relies on local context while only a few instances of long-distance dependencies are present. For example, in Figure~\ref{fig:code} (bottom left), generating assignment operators and parentheses only depend on the current statement, whereas to generate variables such as \texttt{x} and \texttt{conv\_block}, the model needs to look at neighboring statements. Intuitively, we can exploit such locality to sparsify the attention to achieve better efficiency. We further verify this observation by counting the distribution of average attention scores between two tokens within different distances. As shown in Figure~\ref{fig:window_attn}, a large portion of attention is concentrated within a narrow window. Notably, a fixed window of 256 covers more than 90\% of the attention weights. This sparsity enables us to apply a sliding window attention mechanism~\citep{beltagy2020longformer,bigbird}.

Formally, given the linear projections $Q$, $K$, $V$, the self-attention scores in Transformer are calculated as follows:
\begin{equation}
\label{equ:attn}
\mathit{Attention}(Q,K,V) = \operatorname{softmax}(\frac{QK^T}{\sqrt{d_k}}+M)V
\end{equation}
where $M$ is a mask matrix (to be completed in Equation~\ref{equ:finalm}) to control the context a token can attend to when computing its contextual representation. If the $i$-th token is allowed to attend to the $j$-th token, then $M_{ij}$ is set to 0, otherwise  $-\infty$.

For window attention, the mask attention matrix $M^\mathit{window}$ is calculated as follows:
\begin{equation}
    M^\mathit{window}_{ij} = 
    \begin{cases}
        0 & \text{if}\ i-j \leq  w \\
        -\infty & \text{otherwise}
    \end{cases}
\end{equation}

 where $w$ is the window size. This window attention pattern reduces the complexity of the self-attention mechanism by limiting the receptive field size of each token to a small window of size $w$ at each layer. The computation complexity of this pattern is $O(n \times w)$, which scales linearly with input sequence length $n$. After applying $N$ transformer layers of such sliding window attention, the receptive field size increases to $N\times w$ at the top layer. Since each token only attends to $w$ tokens to its left rather than the entire preceding sequence, the model can achieve faster inference speed.

\subsection{Bridge Attention}
Window attention is good at handling local dependencies and it also has a wide receptive field as discussed above.
However, if a token needs to access tokens from a distance of $L$ tokens away, it would require $\lceil \frac{L}{w} \rceil$ hops through window attention. This makes it challenging to access information from distant context as the attention score between them will be greatly reduced due to accumulation through multiplication.
Thus, we introduce a new type of special token, namely \textit{bridge tokens}, to aggregate local information for global access. Bridge tokens can attend to a fixed length of tokens and be attended from all subsequent tokens. From the perspective of representation learning,  a bridge token can be seen as a learned representation for the corresponding slice of code. 

Specifically, we insert $m$ bridge tokens $S_b$ every $\lceil \frac{n}{m} \rceil$ tokens and use a separate set of projections, $Q_b$ , $K_b$ , $V_b$ to compute attention scores for the bridge attention. The bridge tokens do not involve next token prediction but they are used to aggregate information from the preceding $\lceil \frac{n}{m} \rceil$ tokens. The use of additional projections allows for the ability to model different types of attention.
Finally, the mask matrix for bridge attention is calculated as follows:
\begin{equation}
    M^\mathit{bridge}_{ij} = 
    \begin{cases}
        0 & \text{if}\ j\in S_b \ \text{and} \  i\geq j  \\
        0 & \text{if}\ i\in S_b \ \text{and} \ i-j \leq \lceil \frac{n}{m} \rceil \\
        -\infty & \text{otherwise}
    \end{cases}
\end{equation}
The complexity of bridge attention is $O(m \times n) \approx O(n)$ where $m\ll n$. %
Compared to stacked window attention, bridge attention allows each token to attend to any preceding token with at most 2 hops, which enables the model to effectively access long-range context.

\begin{figure}[t!]
    \centering
    \includegraphics[width=0.85\columnwidth]{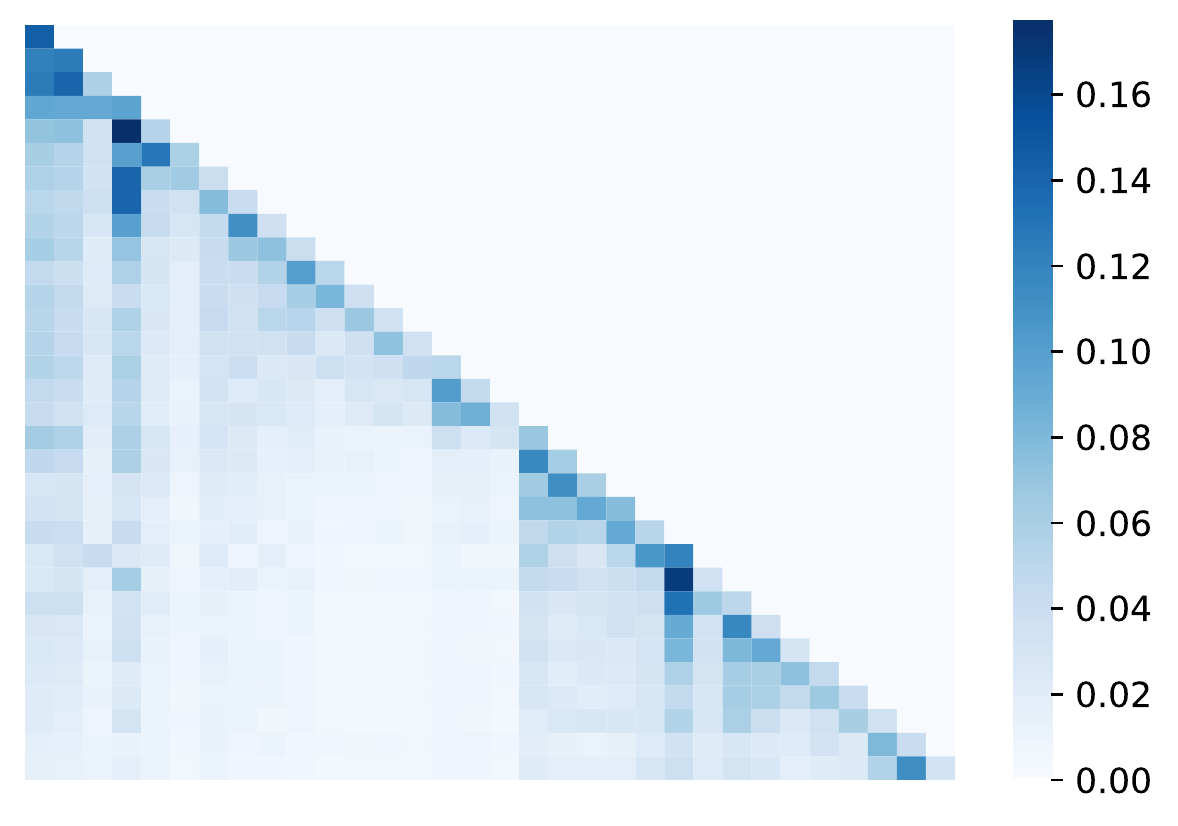}
    \caption{Visualization of the attention matrix (partially shown for clarity) in CodeGPT~\citep{lu2021codexglue}. The attention matrix is averaged across all Transformer layers.}
    \label{fig:memory_attn}
\end{figure}
\subsection{Global Attention}
Identifiers with the global scope, for example, package imports, global functions, classes and their member functions (\ie methods), can be called from any location within a file. For long code, a local sliding window cannot capture such information that should be globally accessible. This is especially outstanding for package imports, which are usually located at the beginning of a file. For example, in Figure~\ref{fig:code}, without knowing the user has imported \texttt{torch.nn.functional} as a new identifier \texttt{F}, the model cannot be sure whether to use \texttt{F} or the full package name.\footnote{Both are common code styles in PyTorch.} Also, to call the forward function of \texttt{conv\_block1} and \texttt{conv\_block2}, the model needs access to the original definition of the class \texttt{ConvBlock}, which also falls outside the current sliding window.
Directly accessing these tokens is similar to how human programmers quickly refer to definitions in the code.
This global effect can also be observed in the visualization of the attention matrix in CodeGPT~\citep{lu2021codexglue}. As shown in Figure~\ref{fig:memory_attn}, some tokens seem to have a global impact while others only matter locally. 

Therefore, in addition to bridge tokens, where the model automatically learns to aggregate globally useful information, we add another type of global token, namely \textit{memory tokens}, to inject code heuristics to the attention. Specifically, we leverage the structure of code with \textit{tree-sitter} to parse the code into an abstract syntax tree (AST). Then, we find all statements that include a package import, class or function definition and grant the line feeds (LF, \verb$\n$) of those statements global access. We denote the set of positions of these line feeds as $G$, where $k=|G|$. The mask matrix $M^{global}$ of global attention is calculated as follows:
\begin{equation}
    M^\mathit{global}_{ij} = 
    \begin{cases}
        0 & \text{if}\ j\ \in G \ \text{and} \  i\geq j  \\
        -\infty & \text{otherwise}
    \end{cases}
\end{equation}
The complexity of the global attention is $O(kn) \approx O(n)$, as we have $k \ll n$, where $n$ is the length of the sequence.

Unlike previous work \cite{clement2021long} which extracts the most significant statements and encode them in a fixed-length context window, our global attention requires less memory and can reuse previously encoded hidden states.

Finally, considering all three types of attention together, $M$ in Equation~\ref{equ:attn} becomes:
\begin{equation}
\label{equ:finalm}
    M = \max(M^\textit{window}, M^\textit{bridge}, M^\textit{global})
\end{equation}
where $\max$ is the element-wise maximum function.

\begin{table*}[t]
\caption{Experimental results on the Long Code Completion (LCC) dataset.}
\label{tab:longcompletion}
\begin{center}
\begin{tabular}{lccccccccc}
    \toprule
    \multirow{2}{*}{Model} & \multirow{2}{*}{\#Param.} & \multirow{2}{*}{Memory} & \multirow{2}{*}{Runtime} & \multicolumn{2}{c}{Python} & \multicolumn{2}{c}{Java}& \multicolumn{2}{c}{C\#} \\
    \cmidrule(lr){5-6} \cmidrule(lr){7-8} \cmidrule(lr){9-10}
    & & & & EM &Edit Sim&EM &Edit Sim&EM &Edit Sim  \\
    \midrule
    OpenAI Codex & 12B&-&-& 39.65& 68.97& 43.15 &72.05 & 53.89&77.93\\
    \midrule
    Transformer & 124M&191M&750ms& 10.64 & 43.64& 15.32 & 47.52& 19.16 & 48.87\\
    GPT-2 & 124M& 191M&750ms &  11.20 & 42.62& 17.09 & 47.18& 20.27 & 48.27\\
    CodeGPT & 124M& 191M&750ms & 12.24 & 43.81& 19.20 & 49.50& 22.58 & 51.03\\
    UniXcoder & 126M& 191M&750ms & 16.55 & 50.22& 23.93 & 55.38& 27.97 & 57.29\\
    \midrule
    LongFormer &150M & 381M& 781ms& 16.79 & 51.07& 24.80 & 56.03& 29.75 & 58.23\\
    BigBird & 128M& 205M& 804ms&17.03  &51.14 &25.19 &56.91&30.27&58.66\\
    \midrule
    LongCoder & 150M& 211M& 812ms& \bf{17.88} & \bf{55.07}& \bf{26.42} & \bf{61.21}& \bf{31.34} & \bf{64.37}\\
    \ - w/o pretrain &150M & 211M& 812ms& 17.61 &54.82 &25.96 & 60.81 &31.22 &64.18\\
    
    \bottomrule
\end{tabular}
\end{center}

\end{table*}

\section{Experiments}
\subsection{Experimental Settings}
\paragraph{Baselines}
We evaluate LongCoder against several publicly available pre-trained %
code generation models, including GPT-2~\cite{radford2019language}, CodeGPT~\cite{lu2021codexglue}, and UniXcoder~\cite{guo2022unixcoder}. GPT-2 is pre-trained on a text corpus and CodeGPT is pre-trained on the CodeSearchNet dataset~\cite{husain2019codesearchnet} using next token prediction as the objective. UniXcoder based on UniLM~\citep{unilm} is pre-trained on a cross-modal dataset that includes code, text, and abstract syntax trees. Additionally, we also compare LongCoder with sparse Transformer models, such as LongFormer~\cite{beltagy2020longformer} and BigBird~\cite{bigbird}. LongFormer uses  a dilated sliding window to model long sequences in the generation task, while BigBird has an attention pattern that includes random, window, and global attention. In addition to these comparable baselines, we also report the performance of OpenAI Codex on LCC for reference. Note that Codex is 100$\times$ larger than other models and is likely to have seen the test set of LCC in its pretraining thus is not directly comparable.
\paragraph{Benchmarks}
We evaluate the performance of LongCoder and the baselines on two benchmarks: LCC (introduced in Section~\ref{sec:lcc}), and the code completion task benchmark in CodeXGLUE~\cite{lu2021codexglue}. CodeXGLUE provides PY150~\cite{raychev2016probabilistic} and JavaCorpus~\cite{allamanis2013mining} datasets in Python and Java for line-level code completion. The statistics for the context length of the CodeXGLUE test datasets are listed in Table \ref{codexglue-sta}. We can see that the context length of the input sequence is 5 times shorter than LCC, and only a small portion of the samples require modeling for long code sequences. The objective of evaluating the performance of sparse models on the CodeXGLUE dataset is to examine their effectiveness in scenarios where the code context is relatively short.
 \begin{table}[t]
\caption{Data statistics about the context length of CodeXGLUE test dataset. 25\%/50\%/75\% refer to the first/second/third quartile.}
\label{codexglue-sta}
\begin{center}
\begin{tabular}{lcccc}
    \toprule
    Language&Average &25\% &50\% &75\% \\
    \midrule
    Python &477.8& 83& 197& 502\\
    Java  &365.0& 74& 171& 397\\
    \bottomrule
\end{tabular}
\end{center}
\end{table}

Moreover, longer context can benefit applications including cross-file code completion~\citep{repobench}. We test the performance of LongCoder on the cross-file-random (XF-R) setting of RepoBench~\citep{repobench}. The task is to predict the next line of code based on a given in-file context, consisting of import statements and preceding lines before the target line, as well as a cross-file context, comprising snippets from other files in the code repository, parsed by import statements.

\paragraph{Evaluation Metrics} We report the number of parameters, inference memory consumption, runtime, Exact Match (EM) and Edit Similarity (Edit Sim) of the baselines. The inference memory consumption and runtime per example are calculated using a beam search with  beam size of 5 and maximum generation length of 64 on a single V100 GPU.

\begin{table}[t]
\caption{Results on CodeXGLUE code completion benchmark.}
\label{table:codexglue}
\begin{center}
\begin{small}
\begin{tabular}{lcccc}
    \toprule
    \multirow{2}{*}{Model} & \multicolumn{2}{c}{PY150} & \multicolumn{2}{c}{JavaCorpus} \\
    \cmidrule(lr){2-3} \cmidrule(lr){4-5}
    &EM &Edit Sim&EM &Edit Sim  \\
    \midrule
    Transformer & 38.51 & 69.01& 17.00 & 50.23\\
    GPT-2 &41.73 & 70.60&27.50 & 60.36\\
    CodeGPT & 42.37 & 71.59& 30.60 & 63.45\\
    UniXcoder& 43.12&72.00& 32.90&65.78\\
    \midrule
    LongCoder & \bf{43.77}&\bf{73.37}& \bf{33.13}&\bf{67.32}\\
    \bottomrule
\end{tabular}
\end{small}
\end{center}
\end{table}

\begin{table}[t]
\caption{Cross-file code completion results on RepoBench XF-R~\citep{repobench}.}
\label{table:cross_file}
\begin{center}
\begin{small}
\begin{tabular}{lcccc}
    \toprule
    \multirow{2}{*}{Model} & \multicolumn{2}{c}{Python} & \multicolumn{2}{c}{Java} \\
    \cmidrule(lr){2-3} \cmidrule(lr){4-5}
    &EM &Edit Sim&EM &Edit Sim  \\
    \midrule
    Transformer &7.0 & 38.3&5.8 & 34.4\\
    GPT-2 &15.5 & 48.2&11.6 & 41.4\\
    CodeGPT &16.6 & 49.1& 13.3 & 44.2\\
    UniXcoder& 18.0&53.4& 16.5&50.0\\
    \midrule
    LongCoder & \bf{21.4}&\bf{59.7}& \bf{19.7}&\bf{60.1}\\
    \bottomrule
\end{tabular}
\end{small}
\end{center}
\end{table}

\begin{table*}[t]
\caption{Ablation study on LongCoder without pre-training. }
\label{tab:longcompletion_ablation}
\begin{center}
\begin{tabular}{lcccccc}
    \toprule
    \multirow{2}{*}{Model} &  \multicolumn{2}{c}{Python} & \multicolumn{2}{c}{Java}& \multicolumn{2}{c}{C\#} \\
    \cmidrule(lr){2-3} \cmidrule(lr){4-5} \cmidrule(lr){6-7}
    &EM &Edit Sim&EM &Edit Sim&EM &Edit Sim  \\
    \midrule
    LongCoder&17.61 &54.82 &25.96 & 60.81 &31.22 &64.18\\
    \ w/o memory tokens & 16.80 & 53.90& 24.87 & 60.35& 30.16 & 63.61\\
    \ w/o bridge tokens & 17.54&51.92&25.88&57.48&30.80&59.09\\
    \ w/o out-of-window context & 16.66 & 50.63& 24.17 &55.97& 28.81& 57.79\\
    \ w/ equidistant memory tokens & 17.16 & 54.25& 24.96 & 60.42& 30.31 &63.65\\
    
    \bottomrule
\end{tabular}
\end{center}
    
\end{table*}

\subsection{Training Details}
We set the maximum length of code context to 512 and 4096 for non-sparse and sparse models, respectively. In order to make a fair comparison between sparse models and non-sparse models, we set the window size $w$ to 512 so that both types of models maintain the same local context length during inference. Note that this setting is different from the original setting of RepoBench~\citep{repobench}, thus the results are not directly comparable to those reported in \citet{repobench}. For sparse models, we use the parameters of  UniXcoder released by \citet{guo2022unixcoder} to initialize the models. For LongCoder, we set the maximum size of bridge tokens $n$ and global tokens $k$ as 16 and 64, respectively. To ensure fair comparison with other models, we pre-train LongCoder on the CodeSearchNet dataset using the same next token prediction objective and pre-training setting as baselines \cite{lu2021codexglue,guo2022unixcoder}. During fine-tuning, we use the Adam optimizer with a batch size of 16 and a learning rate of 2e-4. We fine-tune the model for 10 epochs and perform early stopping on the development set. Note that although the maximum context sequence length is 4096, during inference, we only retain a cache of at most 592 tokens for past key and value hidden states to maintain efficiency in terms of computational resources. 

\subsection{Experimental Results}
Table~\ref{tab:longcompletion} illustrates the comparison results of LongCoder with other models on the LCC dataset. The results reveal that the sparse models (i.e., the last two groups) have superior performance compared to the non-sparse models (i.e., the second group) on both EM and Edit Sim metrics, and they also maintain a similar inference speed. LongFormer is initialized using the parameters of UniXcoder, with the sole difference being the use of a sliding window attention mechanism. This mechanism allows the model to maintain a consistent inference speed while having a larger receptive field, resulting in improved performance. This demonstrates the effectiveness of the sliding window attention mechanism in code completion tasks. Compared to other sparse models, LongCoder achieves an improvement of 0.8\%--1.3\% in Exact Match score and 4.0\%--6.0\% in Edit Similarity, which reveal the effectiveness of our proposed bridge and global attention mechanisms.
Table \ref{table:codexglue} shows the result of LongCoder on CodeXGLUE code completion benchmarks. It can be observed that LongCoder achieves state-of-the-art performance, which illustrates its effectiveness in scenarios where the code context is short. As shown in Table~\ref{table:cross_file}, LongCoder has an even larger advantage compared to UniXcoder, indicating its potential in more complex scenarios.

\subsection{Ablation Study}
To better understand the impact of different components on  overall performance, we conduct an ablation study on LongCoder, and the results are shown in Table \ref{tab:longcompletion_ablation}.
We can see that the average score of Exact Match drops by approximately 1\% when memory tokens are removed (\textbf{w/o memory tokens}), which demonstrates the importance of these tokens. On the other hand, when bridge tokens are removed (\textbf{w/o bridge tokens}), the average score drops by about 3\% in terms of Edit Similarity. This is likely because bridge tokens assist LongCoder in understanding the semantics of the code context and generating more accurate patterns, while memory tokens enable it to access concrete identifiers with global scope, thus improving accuracy on libraries, classes, and functions invoked. 
Additionally, we observe that selecting one as a memory token every 64 tokens (\textbf{equidistant memory tokens}) results in worse performance than LongCoder, indicating that the advantage of the memory tokens is not solely due to increased context length. We also evaluate the performance of LongCoder by only using code context within the window size during inference to verify if the improvement is solely attributed to the use of long code context or whether other factors such as fine-tuning settings also contribute. By only keeping the last 512 tokens as code context (\textbf{w/o out-of-window context}), we can see that the performance is nearly the same as UniXcoder in Table \ref{tab:longcompletion}, which shows the importance of modeling long code context.

\begin{figure*}[t!]
    \centering
    \includegraphics[width=0.99\linewidth]{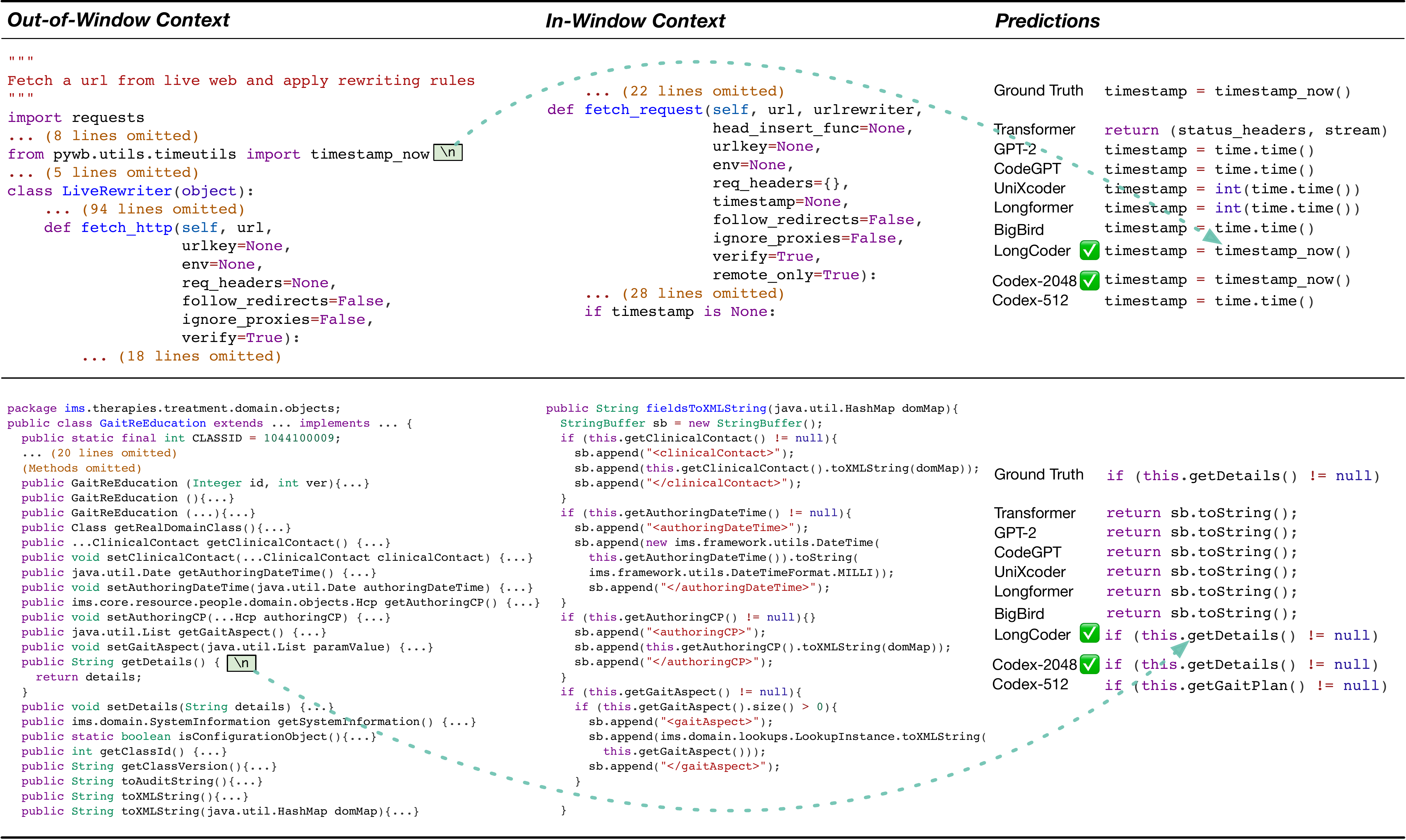}
    \vskip -0.6em
    \caption{Two LCC examples of Python \textbf{\textit{(top)}} and Java \textbf{\textit{(bottom)}} code and predictions of different models. Codex-2048 refers to the original Codex model with the maximum context length of 2,048 while Codex-512 is the same model with a maximum context length set to 512. Key information is highlighted with arrows.}
    \label{fig:case}

\end{figure*}
\subsection{Case Study}
We also conduct a case study to demonstrate the effectiveness of LongCoder, as shown in Figure \ref{fig:case}. We provide two examples in the Python and Java programming languages and output predictions from different models. 
(1) From the Python example, we can see that all models infer the correct intended outcome, which is to assign the current timestamp to the \texttt{timestamp} variable. However, only LongCoder and Codex-2048 produce the correct result. This is primarily because these two models are able to refer to the import statement at the beginning of the file, which imports the \texttt{timestamp\_now} function. Codex-2048 uses a long context window to cover the entire file, but this approach increases memory consumption and decreases inference speed as discussed above. Additionally, as the failure of Codex-512 shows, even a large powerful model can struggle to identify the correct function from other candidates if the required information exceeds the window size. In contrast, LongCoder utilizes a more efficient memory attention mechanism, storing information based on the scope of different statements. This method is more effective, allowing access to statements from the global scope while remaining efficient.
(2) In the Java example, the function to be completed aims to convert a \texttt{HashMap} variable into an XML string. The function sequentially calls the getter functions of the \texttt{GaitReEducation} class and has already completed calling the \texttt{getGaitAspect} function. From the out-window context, it is clear that the next call should be made to the \texttt{getDetails} function. In order to correctly complete the function, it is essential to keep track of all function definitions. As seen in the output results, only LongCoder and Codex-2048, which both make use of long code context, can predict the correct results. Additionally, it can be observed that Codex-512, due to its limited context, can only make a guess for a member function. We can see that LongCoder leverages the structure of the code to analyze the scope of statements and stores those that have potential long-term dependencies. This not only improves performance but also achieves comparable efficiency in terms of computational resources during inference.

\section{Discussion}
\paragraph{Limitations} One limitation of LongCoder is its small size. Due to resource constraints, we are not able to train a large model that is comparable to Codex. Besides, to compare fairly with other baselines, we only pre-train LongCoder on a small-scale corpus (CodeSearchNet). It would be interesting to see how the idea of sparse attention scales with more data.

Another limitation of our work is the evaluation datasets. Many existing code datasets and LCC share the same source of data from GitHub. The same data can appear in the pretraining data, making the evaluation less reliable. Additionally, models with larger-scale pretraining are even more likely to have seen the test data before. For example, OpenAI Codex is trained on all GitHub repositories, that undoubtedly, include most (if not all) test data in PY150, JavaCorpus, and LCC. As the code completion models have seen wider adoption in software development,  future evaluation can be ``self-fulfilling''. For example, GitHub CoPilot\footnote{\url{https://github.com/features/copilot}} is a popular commercial code completion tool powered by OpenAI Codex. A lot of code generated by Codex may have already been submitted to GitHub. This could give widely-used models like Codex an advantage if it is evaluated on a dataset with a data source of the latest GitHub repositories, as we could be evaluating Codex on its own input. To address these problems, we need the community to contribute new, clean, and high-quality datasets with code from private projects to support future research.

\paragraph{Future Work} LongCoder opens up new research opportunities in code generation not only within a large file, but also across multiple files. For example, we could allow the model to look at other files in the project for even more accurate code completion. It could enable new applications including automatically extracting package requirements, generating build files, refactoring the project, etc. 

\subsubsection*{Acknowledgments}
Jian Yin is the corresponding author. Daya Guo and Jian Yin are supported by the National Natural Science Foundation of China (U1911203, U2001211, U22B2060), Guangdong Basic and Applied Basic Research Foundation (2019B1515130001), Key-Area Research and Development Program of Guangdong Province (2020B0101100001). We would like to thank the anonymous reviewers and the meta-reviewer for their insightful comments.

\bibliography{main}
\bibliographystyle{icml2023}

\end{document}